\documentclass[prd,showpacs,nofootinbib,preprintnumbers]{revtex4}

\usepackage{graphicx}
\usepackage{amssymb}
\usepackage{pifont}

\def\lQ{\Lambda_{\rm QCD}}
\newcommand{\be}{
\begin{equation}
}{\bf }
\newcommand{\ee}{
\end{equation}
}
\newcommand{\bea}{
\begin{eqnarray}
}
\newcommand{\eea}{
\end{eqnarray}
}
\def\als{\alpha_{\rm s}}
\def\siml{{\ \lower-1.2pt\vbox{\hbox{\rlap{$<$}\lower6pt\vbox{\hbox{$\sim$}}}}\ }}

\begin{document}

\title{Inclusive radiative decays of charmonium}
\author{Xavier \surname{Garcia i Tormo}}
\affiliation{High Energy Physics Division, Argonne National Laboratory \\  
              9700 South Cass Avenue, Argonne, IL 60439, USA}
\author{Joan Soto}
\affiliation{Departament d'Estructura i Constituents de la Mat\`eria, Universitat de Barcelona\\
Diagonal 647, E-08028 Barcelona, Catalonia, Spain}
\preprint{ANL-HEP-PR-06-89 ~~ UB-ECM-PF 06/42}
\pacs{13.20.Gd, 12.38.Cy, 12.39.St}

\begin{abstract}
We discuss the theoretical status of inclusive radiative decays of charmonium, with a particular emphasis on the QCD description of the photon spectrum, where progress has occurred in recent years. We also comment on the possible extraction of $\als (M_{J/\psi})$ and on the possibility to gain important information on the nature of $J/\psi$ and $\psi (2S)$.

\end{abstract}

\maketitle

\section{Introduction}

Inclusive radiative decays of heavy quarkonium systems to light hadrons have been a subject of investigation since the
early days of QCD \cite{Brodsky:1977du,Koller:1978qg}. It was thought for some time that a reliable extraction of $\als$
was possible from the inclusive $\gamma gg$ decay normalized to the inclusive $ggg$ decay. However, when the experimental data became available for $J/\psi$ \cite{Scharre:1980yn}, it turned out that the photon spectrum, and in particular the upper
end-point region of it, 
appeared to be poorly described by the theory. The situation was slightly better for the $\Upsilon (1S)$ \cite{Nemati:1996xy}, where, at least, good agreement with QCD was found in the central region \cite{Wolf:2000pm}. In fact, the whole photon spectrum for the $\Upsilon (1S)$ is now well understood thanks to a number of theoretical advances which have taken place in recent years (see \cite{GarciaiTormo:2005ch} and references therein). Here we will mainly translate in a critical way these advances to the case of the $J/\psi$.  

We will stay in the effective theory framework of Non-Relativistic QCD (NRQCD) \cite{Caswell:1985ui,Bodwin:1994jh} and Potential NRQCD (pNRQCD) \cite{Pineda:1997bj,Brambilla:1999xf}, and our terminology will follow that of \cite{Brambilla:2004jw}. Let us remind the reader that heavy quarkonium systems enjoy the hierarchies of scales $m\gg mv\gg mv^2$ and $m \gg \lQ$, where $m$ is the heavy quark mass, $v\ll 1$ the relative velocity of the heavy quarks and $\lQ$ a typical hadronic scale. States fulfilling $\lQ \lesssim mv^2$ are said to be in the weak coupling regime (the binding is essentially due to a Coulomb-like potential) whereas states fulfilling $\lQ \gg mv^2$ are said to be in the strong Coupling regime (the binding is due to a confining potential). States below the open flavor threshold and not too deep are expected to be in the strong coupling regime whereas deep states are expected to be in the weak coupling one. States above (or very close to) the open flavor threshold are not expected to be in either regime.

\section{The photon spectrum}

The contributions to the decay width can be split into direct ($^{dir}$) and fragmentation ($^{frag}$)

\be
\frac{d\Gamma}{dz}=\frac{d\Gamma^{dir}}{dz}+\frac{d\Gamma^{frag}}{dz}
\ee
We will call direct contributions to those in which the observed photon is emitted from the heavy quarks and fragmentation contributions to those in which it is emitted from the decay products (light quarks). This splitting is correct at the order we are working but should be refined at higher orders.
$z\in [0,1]$ is defined as $z=2E_\gamma /M$, $M$ being the heavy quarkonium mass and $E_\gamma$ the photon energy in the heavy quarkonium rest frame.

\subsection{Direct Contributions}\label{secdirect}
The starting point is the QCD formula \cite{Rothstein:1997ac}
\begin{equation}
{d \Gamma\over dz}=z{M\over 16\pi^2} {\rm Im} T(z)\quad \quad T(z)=-i\int d^4 x e^{-iq\cdot x}\left<
V_Q (nS)
 \vert T\{ J_{\mu} (x) J_{\nu} (0)\} \vert
V_Q (nS)
 \right> \eta^{\mu\nu}_{\perp}
\label{gdz}
\end{equation}
where $J_{\mu} (x)$ is the electromagnetic current for heavy quarks in QCD and we have restricted ourselves to $^3S_1$
states.
$q$ is the photon momentum, which in the rest frame of the heavy quarkonium is $q=\left(q_{+},q_{-},
q_{\perp}\right)=(zM,0,0)$,
$q_\pm=q^0\pm q^ 3$. Different approximations to this formula are necessary for the central ($z\sim 0.5$), lower end-point ($z\rightarrow 0$) and upper end-point ($z\rightarrow 1$) regions.   

\subsubsection{The central region}

For $z$ away from the lower and upper end-points ($0$ and $1$ respectively), no further scale is introduced beyond those
inherent of the non-relativistic system.  The integration of the scale $m$ 
in the time ordered 
product of currents in (\ref{gdz}) leads to local NRQCD operators with matching coefficients which depend on $m$ and
$z$. At leading order one obtains
\begin{equation}
\label{LOrate}
\frac1{\Gamma_0} \frac{d\Gamma
_{\rm LO}}{dz} =  
\frac{2-z}{z} + \frac{z(1-z)}{(2-z)^2} + 2\frac{1-z}{z^2}\ln(1-z) - 2\frac{(1-z)^2}{(2-z)^3} \ln(1-z),
\end{equation}
where 
\begin{equation}
\Gamma_0 = \frac{32}{27}\alpha\alpha_s^2e_Q^2
\frac{\langle  V_Q (nS)\vert {\cal O}_1(^3S_1)\vert V_Q (nS)\rangle}{m^2},
\label{gamma0}
\end{equation}
and $e_Q$
is the charge of the heavy quark. The $\alpha_s$ correction to this rate was calculated numerically in
Ref.~\cite{Kramer:1999bf} for the bottomonium case. A reasonable estimate for charmonium maybe obtained by multiplying it by $\als (2m_c)/\als (2m_b)$.
The expression corresponding to (\ref{gamma0}) in pNRQCD is obtained at lowest order 
by just making the substitution 
\begin{eqnarray}
\label{singletWF}
\langle  V_Q (nS) \vert {\cal O}_1(^3S_1) \vert  V_Q (nS) \rangle &=&
 \frac{N_c}{2\pi} |R_{n0}(0)|^2,
\end{eqnarray}
where $R_{n0}(0)$ is the radial wave function at the origin. The final result coincides with the one of the early QCD
calculations \cite{Brodsky:1977du,Koller:1978qg}. The NLO contribution in the 
weak coupling regime reads
\cite{Bodwin:1994jh}, 

\begin{equation}
\label{RelCo}
\frac{d\Gamma _{\rm NLO}}{dz}=C_{\mathbf{1}}'\left(\phantom{}^3S_1\right)\frac{\langle  V_Q (nS)\vert {\cal
P}_1(^3S_1)\vert V_Q (nS)\rangle}{m^4}
\end{equation}
ant it is $v^2$ suppressed with respect to (\ref{LOrate}).
The new matrix element above can be written in
terms of the original one 
\cite{Brambilla:2002nu}

\begin{equation}
\frac{\langle  V_Q (nS)\vert {\cal P}_1(^3S_1)\vert V_Q (nS)\rangle}{m^4}=\left(\frac{M-2m-\mathcal{E}_1/m}{m}\right)\frac{\langle  V_Q
(nS)\vert {\cal O}_1(^3S_1)\vert V_Q (nS)\rangle}{m^2}\left(1+\mathcal{O}\left(v^2\right)\right)
\end{equation}
In the 
weak coupling regime $\mathcal{E}_1/m$ is absent
\cite{Gremm:1997dq}, but in the strong coupling regime it must be kept ($\mathcal{E}_1\sim \lQ^2$ is a bound state independent non-perturbative parameter) .
The matching coefficient can be extracted from an early calculation \cite{Keung:1982jb} (see also \cite{Yusuf:1996av}).
It reads

\begin{equation}
C_{\mathbf{1}}'\left(\phantom{}^3S_1\right)=-\frac{16}{27}\alpha\alpha_s^2e_Q^2\left(\left(F_B(z)+\frac{1}{2} F_W(z)\right)\frac{1}{2}+\frac1{\Gamma_0} \frac{d\Gamma
_{\rm LO}}{dz}\right)\label{rel}
\end{equation}
where 
$F_B(z)$ and $F_W(z)$ are defined in ref. \cite{Yusuf:1996av}\footnote{The last term in (\ref{rel}) was missing in \cite{GarciaiTormo:2005ch}, see footnote 4 of \cite{GarciaiTormo:2006ew}.}.

In the weak coupling regime the contributions of color octet operators start at order $v^4$. Furthermore, 
away of the upper end-point region, the lowest order color octet contribution identically vanishes
\cite{Maltoni:1998nh}. Hence there is no $1/\als$ enhancement in the central region and we can safely neglect these
contributions in this case. However, in the strong coupling regime the color octet contributions may become order $v^2$ and should be kept at NLO.

Then in the weak coupling regime (if we use the counting $\als (m)\sim v^2$, $\als\left(m\als\right)\sim v$) the complete NLO ($v^2$ suppressed)
contribution consists of the $\als$ correction to (\ref{LOrate}), the relativistic corrections in (\ref{RelCo}) and the
corrections to the wave function at the origin up to order 
$\als^2\left(m\als\right)$ \cite{Penin:1998kx,Melnikov:1998ug}.
Using $m=m_c=1.6\,GeV$, $M=3.1\, GeV$, $\als(2m_c)=0.23$ and $\als(m\als)=0.4$ we obtain the solid green curve in Fig. \ref{direpmerg}.

\subsubsection{The lower end-point region}

For $z\rightarrow 0$, the emitted low energy photon can only produce transitions within the non-relativistic bound state
without destroying it. Hence the direct low energy photon emission takes place in two steps: (i) the photon is emitted
(dominantly by dipole electric and magnetic transitions) and (ii) the remaining (off-shell) bound state is annihilated
into light hadrons. For $z$ very close to zero it has a suppression  $\sim z^ 3$ with respect to $\Gamma_0$ (see
\cite{Manohar:2003xv,Voloshin:2003hh} for a recent analysis of this region in QED). 
 Hence, at some point the direct photon emission is overtaken by the fragmentation contributions $\bar Q Q \rightarrow
ggg \rightarrow gg\bar q q \gamma $ \cite{Catani:1994iz,Maltoni:1998nh}. In practise this is expected to happen somewhere between $0.2\lesssim z\lesssim 0.4$, 
namely much before than the $z^3$ behavior of the very low energy direct photon emission can be observed, and hence we shall
neglect the latter in the following. 

\subsubsection{The upper end-point region}

In this region the standard NRQCD factorization is not applicable \cite{Rothstein:1997ac}. This is due to the fact that
small scales induced by the kinematics enter the problem and have an interplay with the bound state dynamics. In order
to study this region, one has to take into account collinear degrees of freedom in addition to those of NRQCD. This can
be done using Soft-Collinear Effective Theory (SCET) \cite{Bauer:2000ew,Bauer:2000yr} as it has been described in \cite{Bauer:2001rh,Fleming:2002sr}. 
This region has only been considered in the weak coupling regime, which we will restrict our discussion to. The color octet contributions are only suppressed by $v^2$ or by $1-z$. Since their matching
coefficients are enhanced by $1/\als (m)$, they become as important as the color singlet contributions if we count $\als
(m)\sim v^2\sim 1-z$. 
The formula one may use for the
semi-inclusive width in the end-point region, which was successful for the bottomonium case, reads   
\be
\frac{d\Gamma^e}{dz}=\frac{d\Gamma^{e}_{CS}}{dz}+\frac{d\Gamma^{e}_{CO}}{dz}
\label{endp}
\ee
where $CS$ and $CO$ stand for color singlet and color octet contributions respectively.

For the color singlet contribution one may use the expression with the Sudakov resummed coefficient in ref.
\cite{Fleming:2004rk}
\[
\frac{1}{\Gamma_0}\frac{d\Gamma^{e}_{CS}}{dz} = \Theta(M-2mz) \frac{8z}9 
\sum_{n \rm{\ odd}} \left\{\frac{1}{f_{5/2}^{(n)}}
\left[ \gamma_+^{(n)} r(\mu_c)^{2 \lambda^{(n)}_+ / \beta_0}  - 
\gamma_-^{(n)} r(\mu_c)^{2 \lambda^{(n)}_- / \beta_0} \right]^2+\right.
\]
\begin{equation}
\label{singres}
\left.+
\frac{3 f_{3/2}^{(n)}}{8[f_{5/2}^{(n)}]^2}\frac{{\gamma^{(n)}_{gq}}^2}{\Delta^2}
\left[ r(\mu_c)^{2 \lambda^{(n)}_+ / \beta_0}  -  
r(\mu_c)^{2 \lambda^{(n)}_- / \beta_0} \right]^2\right\}
\end{equation}
where the definitions for the different functions appearing in (\ref{singres}) can be found in \cite{Fleming:2004rk,GarciaiTormo:2005ch}.

For the color octet contributions we use
\begin{equation}
\frac{d\Gamma_{CO}^{e}}{dz}=\alpha_s\left(\mu_u\right)\alpha_s\left(\mu_h\right)e_Q^2\left(\frac{16M\alpha}{9m^4%
}\right)\int_z^{\frac{M}{2m}}\!\!\! C(x-z) S_{S+P}(x)dx
\end{equation}
$\mu_u\sim mv^2$ and $\mu_h\sim m$ are the ultrasoft and hard scales respectively. 
$C(x-z)$ contains 
the Sudakov resummations of ref. \cite{Bauer:2001rh}. The (tree level) matching coefficients (up to a global factor) and the various shape functions are encoded in $S_{S+P}(x)$. See \cite{GarciaiTormo:2005ch} for a precise definition of these objects.

We would like to comment on the validity of the formulas above. This is
limited by the perturbative treatment of the collinear and ultrasoft gluons. 
For the collinear gluons, entering in the jet functions, we have $1 GeV \lesssim M\sqrt{1-z}$, which for $J/\psi$ implies $z \lesssim 0.9$. The formalism is not reliable beyond that point. For the ultrasoft gluons, entering in the shape functions ($S_{S+P}(x)$), we have $1 GeV \lesssim M(1-z)$, which implies $z \lesssim 0.7$. Hence, due to the latter, we do not really have a reliable QCD description of the upper end-point region for charmonium. However, for the bottomonium system, the shape function above turns out to describe very well the data even in the far end-point region, where it is not supposed to be reliable either. In view of this, we believe that the formulas above may provide a reasonable model for the description of the region $0.7 \lesssim z \lesssim 0.9$. The outcome is the blue dot-dashed curve of Fig. \ref{direpmerg} (the flattening of the curve for $z > 0.85$ is an artifact, see footnote 2 of Ref. \cite{GarciaiTormo:2005ch})

\subsubsection{Merging the central and 
upper end-point regions}\label{subsecmatch}

As we have seen, different approximations are necessary in the central and upper end-point regions. It is then not obvious how the results for the central
and for the upper end-point regions must be combined in order to get a reliable description of the whole spectrum.
When the results of the central region are used in the upper end-point region, one misses certain Sudakov and Coulomb resummations which are necessary because the softer scales $M\sqrt{1-z}$ and $M(1-z)$ become relevant. Conversely, when results for the end-point region are used in the central region, one misses non-trivial functions of $z$, which are approximated by their end-point ($z\sim 1$) behavior. In \cite{GarciaiTormo:2005ch} the following merging formula was proposed, which works reasonably well for bottomonium, 
\be
\frac{1}{\Gamma_0}\frac{d\Gamma^{dir}}{dz}=\frac{1}{\Gamma_0}\frac{d\Gamma^{c}}{dz}+\left(\frac{1}{\Gamma_0
}\frac{d\Gamma^{e}}{dz}-\left.{\frac{1}{\Gamma_0
}\frac{d\Gamma^{e}}{dz}}\right\vert_c\right)
\label{mergingNLO}
\ee
$\vert_c$ means the expansion of the end-point formulas when $z$ approaches the central region. This expansion must be carried out at the same level of accuracy as the one we use for the formulas in the central region. 

\begin{figure}
\centering
\includegraphics[width=7.5cm]{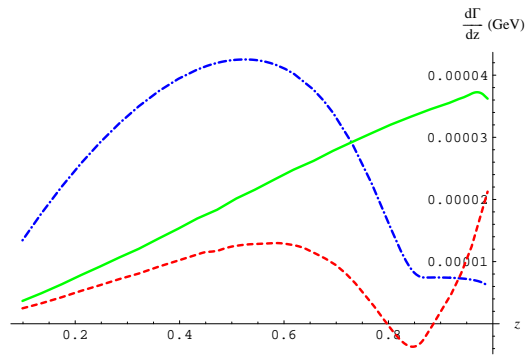}
\caption{Direct contributions in the weak coupling regime. The solid green line corresponds to the calculation for the central region at NLO, which should be reliable up to $z\lesssim  0.7$ . The blue dot-dashed line corresponds to the calculation for the upper end-point region, which is expected to provide a reasonable model for $0.7\lesssim z\lesssim 0.9$. The red dashed line is the curve obtained by merging.}
\label{direpmerg}
\end{figure}

Putting all the ingredients together in formula (\ref{mergingNLO}) we obtain the red dashed line in Fig. \ref{direpmerg} for the direct contributions to the photon spectrum. Note that a deep is generated for $0.8\lesssim z\lesssim 0.9$ which makes the decay width negative. This happens in the region $0.7\lesssim z$ where the calculation of the shape function is not reliable. A deep was also generated in the $\Upsilon (1S)$ case, but the effect was not so dramatic there \cite{GarciaiTormo:2005ch}. We conclude that, unlike in the $\Upsilon (1S)$ case, the limitations in the theoretical description of the end point region make the merging procedure deliver an unsatisfactory description of this end point region for $J/\psi$. Clearly, further work is required to understand better the end-point region, in particular a description assuming $J/\psi$ in the strong coupling regime would be desirable. For the present analysis, one should only use the outcome of the merging procedure for $z < 0.7$, if at all. Indeed, an alternative way to proceed would be to ignore the end-point region and only try to describe data in the central region with the QCD formulas for this region given above.

\subsection{Fragmentation contributions}\label{secfrag}

The fragmentation contributions can be written as
\begin{equation}
\frac{d\Gamma^{\rm frag}}{dz}=\sum_{a = q,\bar q, g} \int_z^1\frac{dx}{x}C_a(x)D_{a\gamma}\left(\frac{z}{x},M\right),
\end{equation}
where $C_a$ represents the partonic kernels and $D_{a\gamma}$ represents the fragmentation functions. The partonic
kernels can again be expanded in powers of $v$ \cite{Maltoni:1998nh}
\begin{equation}
C_a=\sum_{\mathcal{Q}}{\hat C}_a[\mathcal{Q}]{\left< J/\psi \vert\mathcal{Q}\vert J/\psi \right>\over m^{d_\mathcal{Q}-1}}
\end{equation}
where $\mathcal{Q}$ stands for NRQCD operators, ${\hat C}_a[\mathcal{Q}]$ for their matching coefficients, and $d_\mathcal{Q}$ for their dimension. The leading order term in $v$ is the color singlet rate to produce three gluons ($\mathcal{Q}={\cal O}_1(^3S_1)$).
The color octet contributions have a $1/\alpha_s$ enhancement. In the weak coupling regime, which we will assume in the following, they are $v^4$ suppressed, but one should keep in mind that in the strong coupling regime they may become order $v^2$.
Then the color singlet fragmentation contribution is of order $\alpha_s^3D_{g\to\gamma}$ and the color
octet fragmentation are of order $v^4\alpha_s^2D_{g\to\gamma}$ ($\mathcal{Q}=O_8(\phantom{}^1S_0 )$, $O_8(\phantom{}^3P_J )$)
or $v^4\alpha_s^2D_{q\to\gamma}$ ($\mathcal{Q}=O_8(\phantom{}^3S_1 )$). We can use, as before, the counting
$v^2\sim\alpha_s$ to compare the relative importance of the different contributions. The existing models for the
fragmentation functions \cite{Aurenche:1992yc} show us that $D_{q\to\gamma}$ is much larger than $D_{g\to\gamma}$.
This causes the $v^4\alpha_s^2D_{q\to\gamma}$ of the $O_8(\phantom{}^3S_1)$ contribution to dominate in front of the singlet $\alpha_s^3D_{g\to\gamma}$ and the octet $v^4\alpha_s^2D_{g\to\gamma}$ contributions. Moreover, the $\alpha_s$ corrections to the singlet rate  will produce terms of order $\alpha_s^4D_{q\to\gamma}$, that is of the same order as the octet $O_8(\phantom{}^3S_1 )$ contribution, 
which are unknown. This results in a large theoretical uncertainty in the fragmentation contributions, which would be greatly reduced if the leading order calculation of ${\hat C}_q[ O_1(\phantom{}^3S_1 )]$ (this requires a tree level four body decay calculation plus a three body phase space integral) was known.

\begin{figure}
\centering
\includegraphics[width=7.5cm]{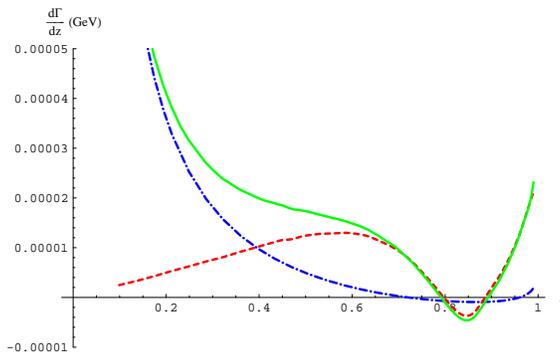}
\caption{The red dashed line corresponds to the direct contributions (merging), the blue dot-dashed line to the fragmentation contributions ($v$ scaling for $O_8(^3S_1)$) and the solid green line is the total.}
\label{mergfragtot}
\end{figure}

\begin{figure}
\centering
\includegraphics[width=7.5cm]{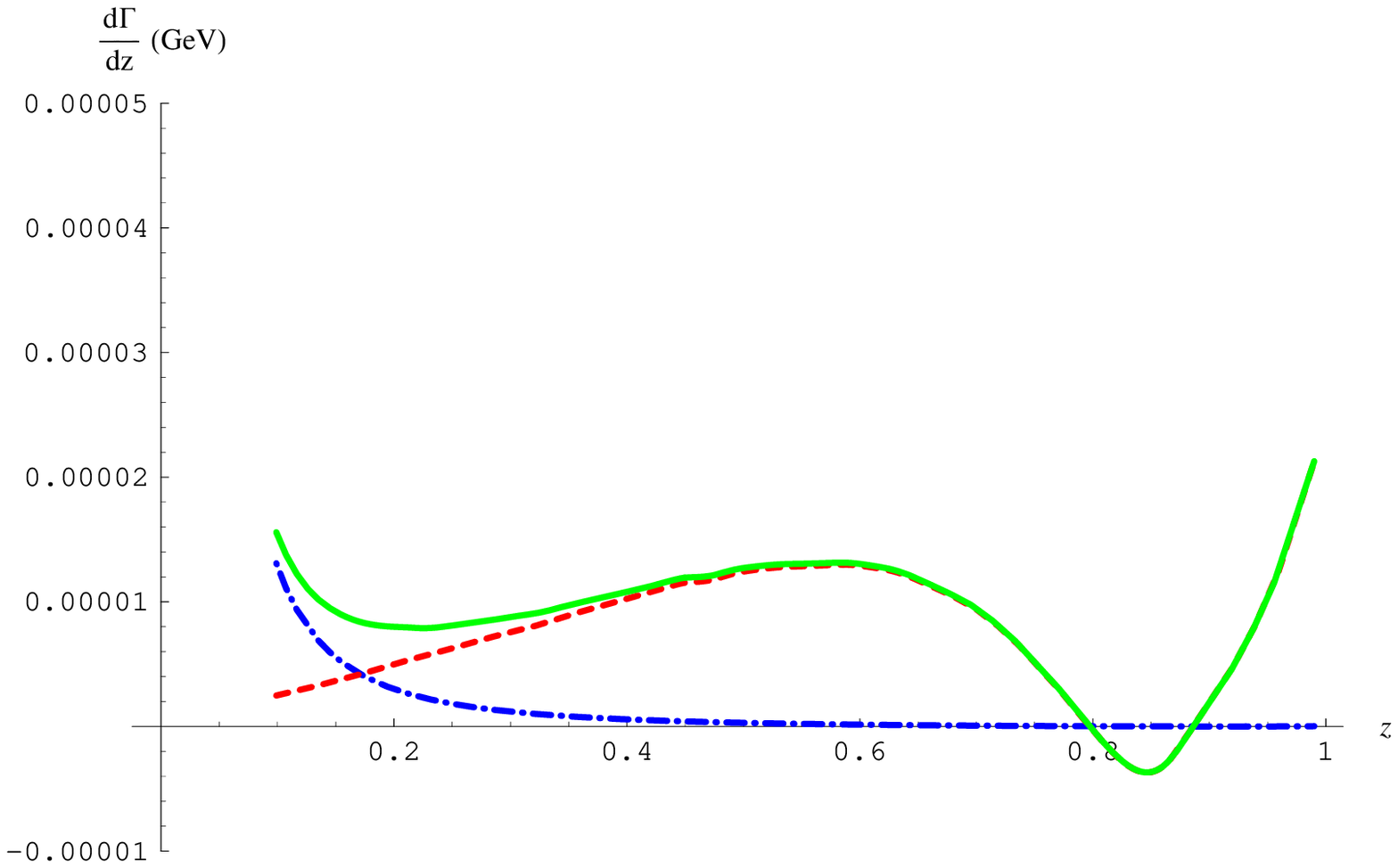}
\caption{The red dashed line corresponds to the direct contributions (merging), the blue dot-dashed line to the fragmentation contributions (lattice, nrqcd algorithm, for $O_8(^3S_1)$) and the solid green line is the total.}
\label{mergfragtot2}
\end{figure}

\begin{figure}
\centering
\includegraphics[width=7.5cm]{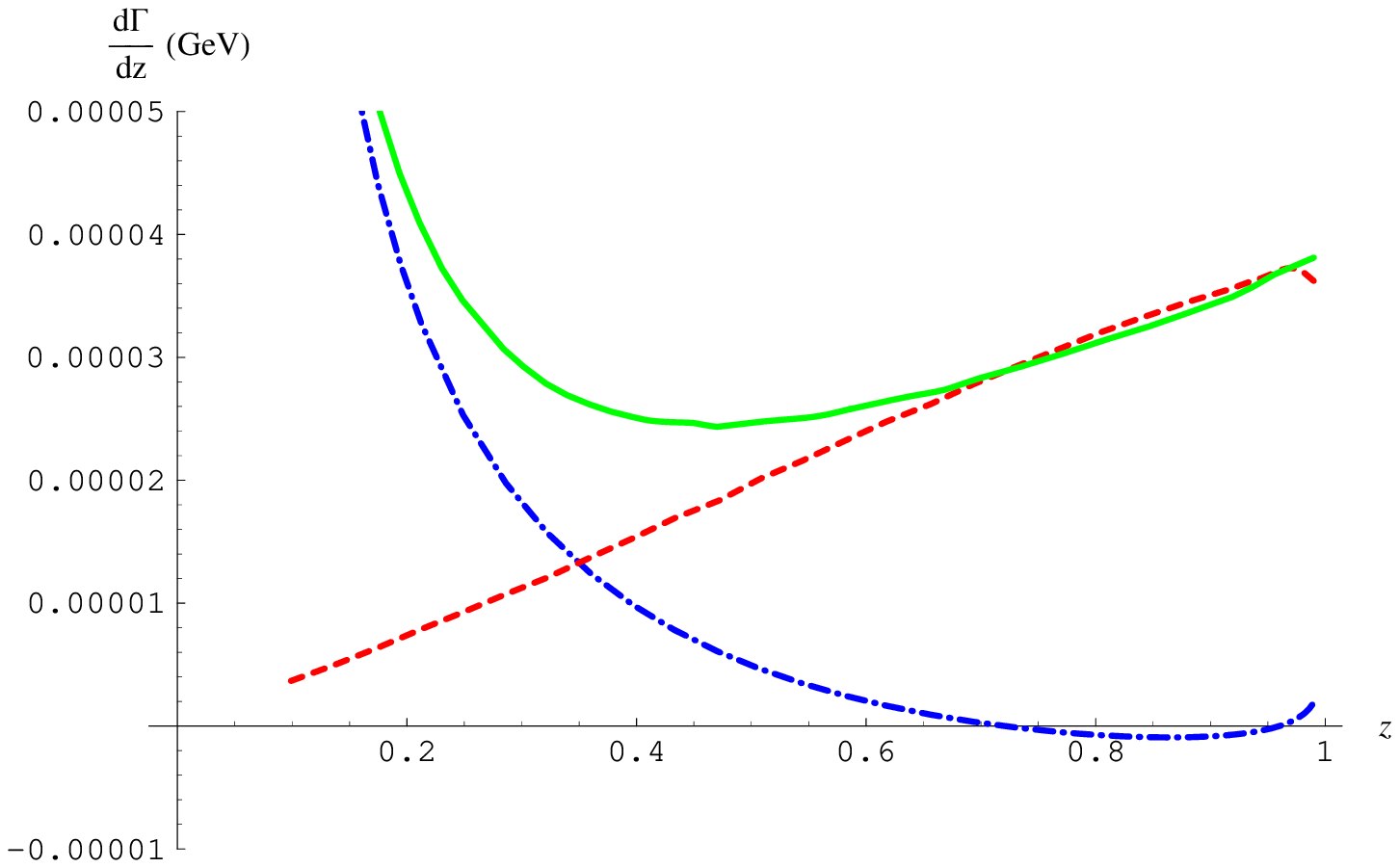}
\caption{The red dashed line corresponds to the direct contributions (central region), the blue dot-dashed line to the fragmentation contributions ($v$ scaling for $O_8(^3S_1)$) and the solid green line is the total.}
\label{centrfragtot}
\end{figure}

\begin{figure}
\centering
\includegraphics[width=7.5cm]{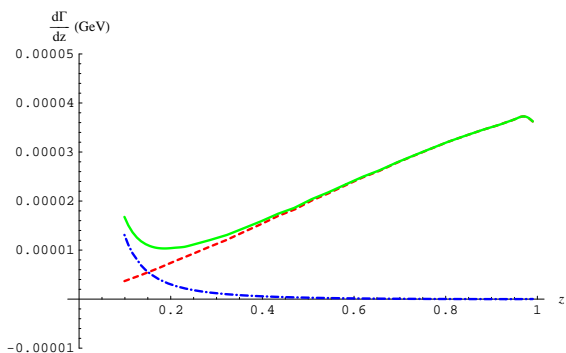}
\caption{The red dashed line corresponds to the direct contributions (central region), the blue dot-dashed line to the fragmentation contributions (lattice, nrqcd algorithm, for $O_8(^3S_1)$) and the solid green line is the total.}
\label{centrfragtot2}
\end{figure}

For the quark fragmentation function we will use the LEP measurement \cite{Buskulic:1995au}  
and for the gluon fragmentation function the model \cite{Owens:1986mp}. These are the same choices as in
\cite{Fleming:2002sr}. For the $O_8 (^1 S_0)$ and $O_8 (^3 P_0)$ matrix elements we will use our estimates in
\cite{GarciaiTormo:2004jw}
\begin{eqnarray}
\left.\left< J/\psi \vert O_8 (^1 S_0) \vert J/\psi \right>\right|_{\mu=M} & \sim & 0.0012\, GeV^3\\
\left.\left< J/\psi \vert O_8 (^3 P_0) \vert J/\psi \right>\right|_{\mu=M} & \sim & 0.0028\, GeV^5
\end{eqnarray}
The estimate for $\left< J/\psi \vert O_8 (^1 S_0) \vert J/\psi \right>$ is compatible with the lattice results \cite{Bodwin:2005gg} (nrqcd and Coulomb algorithms).
For the $O_8 (^3 S_1)$ matrix element the same lattice calculation gives 
\begin{equation}
\left< J/\psi \vert O_8 (^3 S_1) \vert J/\psi \right>_{\textrm{nrqcd}}=0.0005\, GeV^3 \quad \quad \left< J/\psi \vert O_8 (^3 S_1) \vert J/\psi \right>_{\textrm{coulomb}}=0.0002\, GeV^3
\end{equation}
which is much smaller than
using the NRQCD $v$ scaling
\begin{equation}
\left< J/\psi \vert O_8 (^3 S_1) \vert J/\psi \right>\sim v^4\left< J/\psi \vert O_1 (^3 S_1) \vert J/\psi \right>\sim 0.05\, GeV^3
\end{equation}
with $v^2\sim0.3$. With the choices above we obtain the blue dot-dashed curves in Fig. \ref{mergfragtot} ($v$ scaling) and Fig. \ref{mergfragtot2} (lattice, nrqcd algorithm) for the fragmentation contributions. 
These curves turns out to be very sensitive to the value assigned to $\left< J/\psi \vert O_8 (^3 S_1) \vert J/\psi \right>$. 
When we put together direct (red dashed curve in Figs. \ref{mergfragtot} and  \ref{mergfragtot2}) and fragmentation contributions we obtain the solid green curves in Fig.  \ref{mergfragtot} and  \ref{mergfragtot2}, if the merging formula is used, and Fig.  \ref{centrfragtot} and  \ref{centrfragtot2}, if only the central region is taken into account. The shape of this curve in Fig.  \ref{mergfragtot} is in qualitative agreement with the early Mark II results for $0.4\lesssim z \lesssim 0.7 $ \cite{Scharre:1980yn}. 

\section{Extraction of $\als (M_{J/\psi})$}

As mentioned in the introduction, $\als (M_{J/\psi} )$ can in principle be extracted from the ratio $\Gamma (J/\psi \rightarrow \gamma_{\rm direct} X)$ over $\Gamma_{\rm strong} ( J/\psi \rightarrow X)$, $X$ stands for light hadrons, $\gamma_{\rm direct}$ for photons produced from the heavy quarks and $\Gamma_{\rm strong}$ for subtracting from $\Gamma$ the decays mediated by a virtual photon. This maybe done in an analogous way as it has been recently carried out for bottomonia in \cite{Besson:2005jv}. In order to obtain $\Gamma (J/\psi \rightarrow \gamma_{\rm direct} X)$ it is important to have a good QCD description of the photon spectrum since one should be able to disentangle fragmentation contributions from direct ones. This may be done by restricting the fit to data of the QCD expression for direct contributions to the upper end-point region and the part of the central region where fragmentation contributions are negligible. As we have seen in the previous section, we do not have at the moment a good QCD description of the upper end point region for $J/\psi$ and hence a model, like the one in \cite{Field:1983cy}(see also \cite{Field:2001iu}), might be unavoidable. This expression is then used to interpolate data to small $z$ and hence to be able to obtain the full inclusive width for direct photons. Then $\als (M_{J/\psi} )$ may be extracted from 
\bea
R_\gamma &\equiv& \frac{\Gamma(J/\psi \to \gamma_{\rm direct}\,
  X)}{\Gamma_{\rm strong}(J/\psi \to X)} 
= 
\frac{36}{5}\frac{e_c^2\alpha}{\als} \left(1+{\mathcal O}(\als)+{\mathcal O}({v^4\over \als (m)})+\cdots\right)
\eea  
where $e_c^2=4/9$, the ${\mathcal O}(v ^2)$ cancel in the ratio, and $\cdots$ stand for higher order contributions. In the extraction of $\als$ from bottomonium of \cite{Besson:2005jv}, ${\mathcal O}(\als (m) )$ corrections were taken into account but not ${\mathcal O}({v^4\over \als})$ which are of the same order if $\als (m)\sim v^2$ and in practise turn out to be very large. These have been included in \cite{nosaltres}.

\section{Learning about the nature of $J/\psi$ and $\psi (2S)$}

It has recently been shown that if two heavy quarkonium states are in the strong coupling regime then the ratio of their total photon spectrum in the central region is predictable from QCD at NLO \cite{GarciaiTormo:2005bs}. If the spectrum of both $J/\psi$ and $\psi (2S)$ is measured and fits well with the strong coupling regime formula, it would indicate that both $J/\psi$ and $\psi (2S)$ are in the strong coupling regime. Unfortunately, if it does not, we will not be able to learn much, because it may be due to the fact that $J/\psi$ is in the weak coupling regime or to the fact that $\psi (2S)$ is too close to the open flavor threshold for the strong coupling regime to hold for it (or to both).

\section{Conclusions}

A new measurement of the inclusive photon spectrum for radiative $J/\psi$ decays would be of great interest since it has only been measured before by the Mark II collaboration more than 25 years ago. The theoretical progress which has occurred since may allow, among other things, for a sensible extraction of $\als (M_{J/\psi})$. The additional measurement of the photon spectrum for $\psi (2S)$ might shed some light on the nature of these states. No theoretical analysis are available for other states like $\eta_c$s or $\chi_c$s. Experimental measurements would definitively trigger them.

\begin{acknowledgments} 

The work of X.G.T. was supported in part by the U.S. Department of Energy, Division of High Energy Physics, under contract DE-AC02-06CH11357. J.S. acknowledges financial support from MEC (Spain) grant CYT FPA
2004-04582-C02-01, the CIRIT (Catalonia) grant 2005SGR00564, and the RTNs
Euridice HPRN-CT2002-00311 and Flavianet MRTN-CT-2006-035482 (EU).
 
\end{acknowledgments}

\end{document}